\documentclass[pra,twocolumn,showpacs,superscriptaddress]{revtex4-1}
\usepackage{amsthm}
\usepackage{amsmath}
\usepackage{latexsym}
\usepackage{amsfonts}
\usepackage{amssymb}
\usepackage{color}
\usepackage{bbm,dsfont}
\usepackage{graphicx}
\usepackage{hyperref}
\usepackage{subfigure}

\newtheorem{proposition?}{Proposition?}

\theoremstyle{definition}

\newcommand{\ket}[1]{|#1\rangle} %ket
\newcommand{\bra}[1]{\langle#1|} %bra
\newcommand{\id}{\mathbbm{1}} %identity operator
\newcommand{\vx}{\mathbf{x}} %x
\newcommand{\vsigma}{\boldsymbol{\sigma}} %sigma
\newcommand{\nc}{\newcommand}
\nc{\eq}{\begin{equation}}
\nc{\eeq}{\end{equation}}
\nc{\eqa}{\begin{eqnarray}}
\nc{\eeqa}{\end{eqnarray}}
\nc{\ar}{\begin{array}}
\nc{\ear}{\end{array}}
\nc{\bfig}{\begin{figure}}
\nc{\efig}{\end{figure}}
\nc{\dg}{\dagger}
\nc{\eps}{\frac{\epsilon}{2}}
\nc{\juuri}{\sqrt{\Omega^2+(\eps)^2}}
\nc{\sx}{\sigma_x}
\nc{\sy}{\sigma_y}
\nc{\sz}{\sigma_z}
\nc{\spl}{\sigma_+}
\nc{\sm}{\sigma_-}
\nc{\Sx}{\bar{\sigma}_x}
\nc{\Sy}{\bar{\sigma}_y}
\nc{\Sz}{\bar{\sigma}_z}
\nc{\Spl}{\bar{\sigma}_+}
\nc{\Sm}{\bar{\sigma}_-}
\nc{\nn}{\nonumber}
\nc{\noi}{\noindent}
\nc{\omt}{\tilde{\omega}}
\nc{\Somt}{S(\omt)}
\nc{\Somtd}{S^{\dg}(\omt)}
\nc{\got}{\gamma_{\omega}(t)}
\nc{\gmot}{\gamma_{-\omega}(t)}
\nc{\po}{\mathcal{P}}
\nc{\qo}{\mathcal{Q}}
\nc{\adg}{a^{\dg}}
\nc{\gammat}{\tilde{\gamma}}
\nc{\kvec}{\mathbf{k}}

\def\bra#1{\mathinner{\langle{#1}|}}
\def\ket#1{\mathinner{|{#1}\rangle}}

\begin{document}
\title[Dynamical evolution of incompatibility]{Dynamics of incompatibility of quantum measurements in open systems}

\author{Carole Addis}
\email{Ca99@hw.ac.uk}
\affiliation{SUPA, EPS/Physics, Heriot-Watt University, Edinburgh, EH14 4AS, UK}

\author{Teiko Heinosaari}
\email{teiko.heinosaari@utu.fi}
\affiliation{Turku Centre for Quantum Physics, Department of Physics and Astronomy, University of Turku, Finland}

\author{Jukka Kiukas}
\email{jukka.kiukas@aber.ac.uk}
\affiliation{Department of Mathematics, Aberystwyth University, Penglais, Aberystwyth, SY23 3BZ, UK}

\author{Elsi-Mari Laine}
\email{emelai@utu.fi}
\affiliation{Turku Centre for Quantum Physics, Department of Physics and Astronomy, University of Turku, Finland}

\author{Sabrina Maniscalco}
\email{smanis@utu.fi}
\affiliation{Turku Centre for Quantum Physics, Department of Physics and Astronomy, University of Turku, Finland}

\begin{abstract}
The non-classical nature of quantum states, often illustrated using entanglement measures or quantum discord, constitutes a resource for quantum information protocols. However, the non-classicality of a quantum system cannot be seen as a property of the state alone, as the set of available measurements used to extract information on the system is typically restricted. In this work we study how the non-classicality of quantum measurements, quantified via their incompatibility, is influenced by quantum noise and how a non-Markovian environment can be useful for maintaining the measurement resources.
\end{abstract}


\maketitle

%%%%%%%%%%%%%%%%%%%%%%%
\section{Introduction}\label{sec:intro}
%%%%%%%%%%%%%%%%%%%%%%%

The non-classical nature of quantum states is considered to be an essential resource for the emerging quantum technologies.
 This idea is well-established in the context of entanglement theory \cite{Horo09}, and has also been formulated generally in the framework of quantum resource theories \cite{BrGo15,CoFrSp14}.
However, characterising the non-classicality of a quantum system as a property of the state alone has limited practical significance, since the set of available measurements used to extract information on the system is almost always restricted by experimental limitations. 
Accordingly, measurement resources are expected to play a significant role in realistic quantum devices.

The study of measurement resources is motivated also from a fundamental point of view. 
As shown in \cite{We89} there exist entangled states which nevertheless cannot be used to produce quantum correlations in any Bell experiment because of the existence of a specific hidden variable model for the correlations. For certain scenarios based on correlation experiments, the existence of such a classical description is equivalent to joint measurability of suitable observables.
In particular, it has been recently shown that for the CHSH-Bell inequality \cite{WoPeFe09} and EPR-steering scenarios \cite{WiJoDo07,Sk14,UoMoGu14,QuVeBr14}, 
the appropriate quantum resource can be formulated in terms of \emph{incompatibility} of the available local observables. 
It is therefore important to aim for a better understanding of the incompatibility of observables.

Since every real quantum system interacts with its environment \cite{BP07,Da76}, a practical implementation of any quantum protocol has to take the possible effects of noise into account. 
The dynamical behaviour	of nonclassical features of quantum states has been extensively  studied in recent years. 
Specifically, understanding dynamical phenomena such as entanglement sudden death  \cite{ESD1}, or frozen discord  \cite{TimeIND} has been useful for determining the level of isolation from the environment required to implement protocols relying on these quantum features.

In order to investigate the dynamical loss of measurement resources, we need an approach somewhat different from the conventional state-centred view: decoherence and dissipation now take place in the \emph{Heisenberg picture}, rendering the measurements less useful for revealing nonclassical properties. In fact, consider the general setting consisting of preparation of an entangled state shared by Alice and Bob, followed by noisy local evolution (quantum channel) on Alice's side, and local measurements performed by Alice and Bob at the end. In this setting, one would usually write the associated joint probabilities in the Schr\"odinger picture, with the local noise inducing loss of entanglement on the state. However, we can equivalently use the Heisenberg picture, acting on Alice's measurements.

An essential starting point for our investigation is the fact that incompatibility can never be created but is often destroyed by the action of a quantum channel \cite{ibc}. How exactly this ``measurement decoherence'' takes place during actual open system dynamics has not been investigated before and the purpose of this paper is to take the first step in this direction by performing a quantitative analysis of the dynamical evolution of incompatibility. 
We concentrate on two well known microscopic open system models, where the dynamics can be tailored via environment engineering techniques: the phase damping  \cite{PhaseD} and amplitude damping \cite{BP07} evolutions under both Markovian and non-Markovian noise. 
We find that both evolutions exhibit sudden death of incompatibility of quantum measurements, even though no entanglement sudden death occurs. We further study the case of a highly engineered amplitude damping dynamics, given by a photonic band gap environment, which has been found to efficiently protect entanglement in the long time limit \cite{ENTtrap}. We find that even for such a highly engineered scenario the incompatibility of relevant measurements cannot be maintained.
This demonstrates that incompatibility is more fragile than entanglement, which is in line with the recently observed fact that an entanglement breaking channel is incompatibility breaking, but not vice versa \cite{ibc}, \cite{Pusey15}.

The structure of the paper is as follows. In Sec. \ref{incomp} we review general features of quantum incompatibility, as well as the specific quantification introduced in \cite{ourpaper}. 
In Sec. \ref{sec.noise} we motivate the following sections by a simple classical noise model. 
In Sec. \ref{oqs} we proceed to describe incompatibility under quantum dynamical noise, and in Sec. \ref{dyn1} we introduce the microscopic open system models under consideration. 
The numerical results are presented and discussed in Sec. \ref{dyn}, and we conclude the paper in Sec. \ref{conclusions}.

%%%%%%%%%%%%%%%%%%%%%%%%%%%%%
\section{Quantum incompatibility} \label{incomp}
%%%%%%%%%%%%%%%%%%%%%%%%%%%%%
\subsection{Definition of incompatibility and its quantification}

We first recall that a general quantum measurement is described by a POVM, which (assuming a finite set of measurement outcomes) is a map $E:i\mapsto E_i$ assigning a positive operator $E_i$ to each outcome $i$ and satisfying the normalisation condition $\sum_{i} E_i=\id$. 
Two measurements $E$ and $F$ are \emph{compatible} or \emph{jointly measurable} if there exists a measurement $G=(G_{ij})$ such that $E_i=\sum_j G_{ij}$ and $F_j=\sum_i G_{ij}$.
Intuitively, this means that the measurements can be realised together with a single device. 
The definition extends naturally to collections containing more than two measurements. For the purposes of the present investigation, however, studying the evolution pairs is sufficient. If the measurements are not compatible, they are said to be \emph{incompatible}. In this case they can potentially be used for producing nonclassical features as discussed in the introduction.

In order to quantify incompatibility, some of the authors of this paper recently introduced \cite{ourpaper} the concept of a \emph{incompatibility monotone} as a function $\mathcal{I}$ on pairs of observables, required to satisfy the following conditions:
\begin{itemize}
\item[(i)] $\mathcal{I}(E,F)=0$ if and only if $E$ and $F$ are compatible.
\item[(ii)] $\mathcal{I}(\Lambda(E),\Lambda(F))\leq \mathcal{I}(E,F)$ if $\Lambda$ is any channel.
\end{itemize}
The channel $\Lambda$ in (ii) can be either a quantum channel applied before the measurement, or a classical channel applied on the measurement outcomes. This definition of an incompatibility monotone is analogous to entanglement monotones \cite{Horo09}.

In order to define an explicit quantification satisfying the above conditions, one can use \cite{ourpaper} the general resource-theoretic notion of \emph{noise-robustness} \cite{BrGo15,GrPoWi05,AlPiBaToAc07,BaGaGhKa13,erkka,BuHeScSt13}. Basically the idea is to use classical selection noise as a reference. 
In our case, this can be operationally implemented as follows: we think of implementing a measurement $E$ using a noisy device which sometimes (with probability $\lambda$) ignores the actual outcome, and instead outputs randomly an outcome $i$ according to some fixed probability distribution $p=(p_i)$.  The POVM $E^{\lambda, p}$ describing the resulting deformed measurement is then
\begin{equation}\label{deform}
E^{\lambda,p}_i =(1-\lambda)E_i+\lambda p_i\id,
\end{equation}
and for a given pair $(E,F)$ of measurements, we look for the minimal value of $\lambda$ for which the noise-deformed pair $(E^{\lambda,p}, F^{\lambda,p})$ becomes compatible:
\begin{equation}\label{incompmon}
\mathcal{I}_{p}(E,F):=\inf \{\lambda > 0 \mid (E^{\lambda,p},F^{\lambda,p}) \text{ compatible }\}.
\end{equation}
It is easy to see that this function fulfils the properties (i) and (ii) of an incompatibility monotone, and we also know that $\lambda\leq 1/2$ \cite{BuHeScSt13}. There are also other choices for incompatibility monotones, of which we have included a brief discussion in Sec. VII.

%%%%%%%%%%%%%%%%%%%%%%%%%%
\subsection{Incompatibility in a qubit system}\label{incomp-qubit}
%%%%%%%%%%%%%%%%%%%%%%%%%%%

In this paper we focus exclusively on single qubit systems for which the optimisation \eqref{incompmon} can be reduced to solving a single polynomial equation (see discussion in \cite{ourpaper}). In order to describe this, we first need to recall the representation of qubit measurements.

Binary POVMs have only two outcomes, $0$ and $1$. 
Due to the normalisation, we have $E_0=\id-E_1$, so that the operator $0\leq E_1\leq \id$ in fact specifies the POVM completely. In a qubit system we can then write $E_1$ in terms of its Bloch four-vector $x=(x_0,x_1,x_2,x_3)\equiv(x_0,\vx)$:
\begin{equation}
E_1 = \frac 12 (x_0 \id+\vx\cdot \vsigma) \, , 
\end{equation}
where $\vsigma = (\sigma_1,\sigma_2,\sigma_3)$ and $\sigma_j$'s are the usual Pauli matrices. 
The Bloch vector for the other POVM element $E_0$ is given by $x^\perp:=(2-x_0, -\vx)$.
The condition $0\leq E_1\leq \id$ can be compactly expressed by
$x,x^\perp\in \mathcal F_+$,
where
\begin{equation}
\mathcal F_+=\{ x\mid \langle x|x\rangle\geq 0,\, x_0\geq 0\}
\end{equation}
and 
\begin{equation}
\langle x|y\rangle:=x_0y_0-\sum_{i=1}^3 x_iy_i
\end{equation}
is the Minkowski scalar product. To summarise, a binary measurement in a qubit system is specified by a single four-vector $x$ satisfying $x,x^\perp\in \mathcal F_+$.

In general, deciding whether a pair of measurements is compatible requires solving a convex optimisation problem. 
In a qubit system, the problem simplifies considerably \cite{Busch} (see also \cite{SRH08,Yu10}): two measurements, specified by Bloch vectors $x$ and $y$, are compatible if and only if
\begin{equation*}
\mathsf C(x,y)\geq 0 \, , 
\end{equation*}
where
\begin{equation*}
\begin{split}
\mathsf{C}(x,y):=\ & \left[ \langle x|x\rangle\langle x^\perp|x^\perp\rangle\langle y|y\rangle\langle y^\perp |y^\perp\rangle \right]^{1/2}-\langle x|x^\perp\rangle\langle y|y^\perp\rangle \\
+ \ & \langle x|y^\perp\rangle\langle x^\perp| y\rangle+\langle x|y\rangle\langle x^\perp|y^\perp\rangle \, .
\end{split}
\end{equation*}

Given any two measurements with Bloch vectors $x$ and $y$, we can now write down the formula for the noise-robustness-based quantification \eqref{incompmon} of their incompatibility.
Since we only have two outcomes, the classical noise distribution $p$ is written conveniently in terms of the \emph{bias} parameter $-1\leq b \leq 1$:
\begin{equation}
p_1=\frac12(1+b), \quad p_0=\frac12 (1-b) \, .
\end{equation}
Hence, the noise-deformation \eqref{deform} transforms a given Bloch vector $x$ as
$$
x\mapsto N_{\lambda,b}(x):= ((1-\lambda)x_0+\lambda(1+b), (1-\lambda)\vx),
$$
so the incompatibility quantification for a pair $(x,y)$ is given by
\begin{equation}\label{incompmeas}
\mathcal{I}_b(x,y):=\inf\{\lambda>0\mid \mathsf C\left(N_{\lambda,b}(x),\,N_{\lambda,b}(y)\right)\geq 0\} \, .
\end{equation}
Since $\mathcal{I}_b(x,y)\leq 1/2$, and given that $x$ and $y$ are incompatible, we know that $\lambda\mapsto \mathsf C\left(N_{\lambda,b}(x),\,N_{\lambda,b}(y)\right)$ changes sign at exactly one point on the interval $[0,1/2]$, and this point is $\mathcal{I}_b(x,y)$. This value can be easily determined numerically for any given pair $x,y$, as a root of a polynomial equation.

We close this section by recalling an alternative operational interpretation of the number $\mathcal{I}_b(x,y)$ for the case $b=0$. In fact, it coincides with the maximal violation of the CHSH-Bell inequality achievable with Alice's measurements fixed to be $x$ and $y$ \cite{ourpaper}.

%%%%%%%%%%%%%%%%%%%%%%%%%%%%%%%%
\section{Incompatibility under classical noise}\label{sec.noise}
%%%%%%%%%%%%%%%%%%%%%%%%%%%%%%%%

In order to illustrate the relationship between entanglement and incompatibility in noisy systems, it is helpful to begin with a simple case where the noise is purely classical, obtained by mixing with a completely depolarising noise. This is a variant of the selection noise used already in the preceding section to define the incompatibility quantification, and appears in the construction of the well-known isotropic states \cite{Horo09}
\begin{equation}\label{statenoise}
\varrho_s = (1-s) |\Psi_0\rangle\langle\Psi_0| +s \id/d^2 , \quad 0\leq s \leq 1,
\end{equation}
where $\Psi_0$ is a maximally entangled state of a bipartite system. The parameter $t$ describes classical random selection noise added to the maximally entangled state, and one can show that the state loses its entanglement at some critical value $s=s_{\rm entangled}$, remaining separable for $s>s_{\rm entangled}$. However, \emph{the state loses its nonclassicality already before}, in the sense that after some $s=s_{\rm quantum} < s_{\rm entangled}$ a hidden state model \cite{We89,WiJoDo07} exists for any bipartite correlations of the form
\begin{equation}
\mathbb P_s(i,j|a,b)={\rm tr}[\varrho_s A^a_i\otimes B^b_j],
\end{equation}
where $A^1,\ldots,A^n$ and $B^1,\ldots,B^m$ are local measurements accessible to the two parties Alice and Bob, respectively. Hidden state models are specific hidden variable models for which the correlations $\mathbb P_s(i,j|a,b)$ do not violate any Bell inequality. Moreover, the nonexistence of such a model is equivalent to the possibility of EPR-steering \cite{WiJoDo07}, which has practical applications to e.g. one-sided secure Quantum Key Distribution (QKD) and subchannel discrimination \cite{GRTZ02,PiWa14}. 

For a fixed state, the quantum resource for steering can be entirely associated with measurements. In order to see this, we first write
\begin{equation}\label{evolvedstate}
\varrho_s =(\Lambda^*_s\otimes \id)(|\Psi_0\rangle\langle\Psi_0|),
\end{equation}
where the effect of the classical noise is described by a local quantum channel $\Lambda_s = (1-s){\rm Id}+s \id/d$ and $\Lambda_s^*$ denotes the Schr\"dinger picture. Now, since the noise and measurements are both local, it makes more sense to use the Heisenberg picture:
\begin{equation}
\mathbb P_s(i,j|a,b)=\langle \Psi_0|\Lambda_s(A^a_i)\otimes B^b_j\Psi_0\rangle={\rm tr}[\Lambda_s(A^a_i)^\intercal B^b_j].
\end{equation}
Here, the state $\Psi_0$ is maximally entangled, the transpose is taken in the associated basis, and we observe that the effect of noise is entirely captured by its action on Alice's measurements.
One should notice the following important observation \cite{UoMoGu14,QuVeBr14,ibc}: \emph{the nonexistence of the hidden state model for the correlations $\mathbb P(i,j|a,b)$ is equivalent to incompatibility of Alice's noisy local observables}
\begin{equation}
\Lambda_s(A^a), \quad a=1,\ldots,n \, .
\end{equation}
In particular, this means that the observables $\Lambda_s(A^a)$ lose their incompatibility at the amount of noise $s=s_{\rm quantum}$ where the hidden state model comes into existence. For $s<s_{\rm quantum}$, the measurements have some degree of incompatibility, which can be quantified as described in the preceding section. In the next section, we proceed to investigate this loss of incompatibility under more complicated quantum noise channels arising from dynamical interaction with a quantum environment.

%%%%%%%%%%%%%%%%%%%%%%%%%%%%%%%%%
\section{Incompatibility under quantum dynamical noise} \label{oqs}
%%%%%%%%%%%%%%%%%%%%%%%%%%%%%%%%%

Evolution of an open quantum system can be described by a family $(\Lambda_t)$ of quantum channels indexed by the time parameter $t\geq 0$. This map is typically obtained by assuming unitary dynamics $t\mapsto U_t$ on a total system consisting also of an environment, initialised in a given state $\sigma_{\rm env}$. In fact, given that the system is initially (time zero) prepared in a state $\varrho$, and measured at time $t$ via a POVM $(E_j)$, the probability for the outcome $j$ is given by
\begin{equation}
{\rm tr}[\varrho \Lambda_t(E_j)]:={\rm tr}[U_t(\varrho \otimes \sigma_{\rm env})U_t^* (E_j\otimes \id_{\rm env})].
\end{equation}
This determines a family of completely positive unital maps $\Lambda_t$ acting on the observable algebra of the system, describing the evolution in the Heisenberg picture.
That is, $(\Lambda_t(E_j))$ is the POVM describing the measurement $(E_j)$ performed at time $t$, from the point of view of the initial state $\varrho$.

%%%%%%%%%%%%%%%%%%%%%
\subsection{Dynamics of incompatibility}
%%%%%%%%%%%%%%%%%%%%%

The properties of the measurements may crucially change depending on the point in time at which the measurements are performed.
In particular, incompatibility, as quantified by any incompatibility monotone $\mathcal{I}$, becomes a function of time:
\begin{equation}\label{incomptime}
t\mapsto \mathcal{I}(\Lambda_t(E),\Lambda_t(F)).
\end{equation}

As a motivating starting point for our investigation, a few general remarks can be made concerning the case where $t\mapsto\Lambda_t$ is a continuous \emph{semigroup}, i.e.,
\begin{equation}\label{semigroup}
\Lambda_{t+t'}=\Lambda_t(\Lambda_{t'}(\cdot)).
\end{equation}
In this case it follows directly from the general properties (i) and (ii) of an incompatibility monotone that the function \eqref{incomptime} is nonincreasing. Furthermore (assuming finite system dimension), in a generic (ergodic) case there exists a unique stationary state $\varrho_{s}$, such that
\begin{equation}
\lim_{t\rightarrow\infty} \Lambda_t(\cdot)={\rm tr}[\varrho_s(\cdot)]\id,
\end{equation}
which implies that at the limit, every measurement is trivial. Hence one expects that the incompatibility function \eqref{incomptime} will eventually decrease to zero under the assumption \eqref{semigroup}.

The purpose of the present paper is to study the behaviour of the function \eqref{incomptime} for certain concrete choices of the dynamical map $\Lambda_t$ on two-level systems, using the incompatibility monotone \eqref{incompmeas}. More general scenarios could easily be handled with more numerical effort. However, the two-level system already exhibits the relevant phenomena to the extent sufficient for the illustrative purposes of the present paper.
We represent a given dynamical map $t\mapsto \Lambda_t$ in the Heisenberg picture, and then compute the associated evolution $t\mapsto \Lambda_t(E)$ for any given measurement $E$, in terms of its Bloch vector $x(t)$, where $x(0)$ specifies the initial operator $E_1$. 

The dynamical maps considered in this paper act in the Schr\"odinger picture on a matrix $T=[T_{ij}]$ as
\begin{align}
\begin{pmatrix} T_{00} & T_{01}\\ T_{10} & T_{11} \end{pmatrix} \mapsto  \begin{pmatrix} a(t) T_{00} & c(t) T_{01}\\ c(t)^* T_{10} & T_{11} +(1-a(t))T_{00}\end{pmatrix}\label{evomatrix},
\end{align}
where $a(t)$ and $c(t)$ are explicit functions of time $t$.
The Heisenberg representation of this evolution is determined by the equation
\begin{equation}
\text{tr}[\Lambda^*_t(T)S]=\text{tr}[T\Lambda_t(S)] \, , 
\end{equation}
where $T$ and $S$ are arbitrary two-by-two matrices. 
Fixing $S= \frac{1}{2}(x_0\id+\vx\cdot\vsigma)$ corresponding to an initial Bloch vector $x=(x_0,\vx)$, we then find the Bloch vector $x(t)$ of $\Lambda_t(S)$ via
\begin{align}
& x_0(t) = {\rm tr}[\Lambda_t^*(\id) \, S], & x_i(t) & = {\rm tr}[\Lambda_t^*(\sigma_i) \, S],
\end{align}
so that
\begin{align*}
x_0(t) &= x_0 + (a(t)-1)x_3\\
x_1(t) &= {\rm Re}[c(t)]x_1-{\rm Im}[c(t)]x_2\\
x_2(t) &= {\rm Im}[c(t)]x_1+{\rm Re}[c(t)]x_2\\
x_3(t) &= a(t) x_3.
\end{align*}
Given two initial vectors $x$ and $y$, the incompatibility evolution
\begin{equation}
t\mapsto \mathcal I_b(x(t),y(t))
\end{equation}
can then be computed as explained in Subsec. \ref{incomp-qubit}.

Throughout the calculations, if not mentioned otherwise, the initial measurement pair is chosen as the specific maximally incompatible pair with $x_0=y_0=1$, $x_1=y_2=1$, $x_2=y_1=0$ and $x_3=y_3=0$ (denoted $\mathcal{P}_1$). We also study the case $x_0=y_0=1$, $x_1=y_3=1$, $x_2=y_2=0$ and $x_3=y_1=0$ (denoted $\mathcal{P}_2$). Furthermore, we have set the bias parameter $b=0$. More general measurements could be considered in a similar fashion. However, these two cases already exhibit several interesting phenomena, as we will see below.

%%%%%%%%%%%%%%%%%%%%%%%%%%%%
\subsection{Comparison with entanglement dynamics}
%%%%%%%%%%%%%%%%%%%%%%%%%%%%%%

It was shown in section \ref{sec.noise} that incompatibility and entanglement can be compared in a bipartite scenario involving local measurements and a shared state. More specifically, we look at the scenario from the point of view of one party, say, Alice, and compare the incompatibility of her measurements with the entanglement of the shared state. Bob's measurements do not play a role in this setting. We remark that such asymmetry appears naturally in one-sided protocols such as steering \cite{WiJoDo07} mentioned in the Introduction.

Accordingly, we take an ancillary qubit (in addition to our open qubit system described above), and set the initial state of the combined system to be the maximally entangled state $\Psi_0 = \frac{1}{\sqrt{2}}(\ket{00}+ \ket{11})$. We stress that this specific choice is not crucial, since the connection between incompatibility and entanglement holds for any bipartite state \cite{UoBuGuPe14}. Analogous to eq. \eqref{evolvedstate}, we then consider the evolved state
\begin{equation}
\varrho_t = (\Lambda_t^*\otimes {\rm Id})(|\Psi_0\rangle\langle\Psi_0|),
\end{equation}
where $\Lambda_t^*$ is now the local noise channel of the form \eqref{evomatrix}.

According to the discussion in Sec. \ref{sec.noise}, loss of incompatibility of any set of Heisenberg-evolved local observables can now be compared with the loss of entanglement of the state $\varrho_t$.
In particular, we know from the general argument that incompatibility is lost before entanglement. In order to compare finer details of the respective dynamics, we need to choose some quantification also for entanglement of $\varrho_t$, and in our investigation we will use concurrence \cite{Wootters98}; this has the advantage of connecting our work with the extensive existing literature on entanglement dynamics. Moreover, we would like to note that for the models considered here, the entanglement dynamics happens to be equivalent to the dynamics of the information flux \cite{NMPRL} the non-monotonicity of which indicates the non-Markovian character of the dynamics.

A straightforward calculation shows that the matrix of the evolved state $\varrho_t$ has nonzero elements only in the diagonal and antidiagonal entries. 
Thus, the concurrence $\mathcal{E}(\varrho_t)$ has the simple form \cite{Wang}
\eq
\mathcal{E}(\varrho_t)=2 \max\left\{0, K_1(\varrho_t),K_2(\varrho_t)\right\},
\eeq
with 
\begin{align*}
& K_1(\varrho_t)=|\bra{00}\varrho_t\ket{11}|-\sqrt{\bra{10}\varrho_t\ket{10} \bra{01}\varrho_t\ket{01}} \\
& K_2(\varrho_t)=|\bra{01}\varrho_t\ket{10}|-\sqrt{\bra{00}\varrho_t\ket{00} \bra{11}\varrho_t\ket{11}} \,,
\end{align*}
leading to
\begin{equation}
\mathcal{E}(\varrho_t)=|c(t)| \, .
\end{equation} 

%%%%%%%%%%%%%%%%%%%%%%%%%
\section{Description of the dynamics}\label{dyn1}
%%%%%%%%%%%%%%%%%%%%%%%%%

We now describe the two different choices for the dynamical map in detail, including examples of physically relevant scenarios.

%%%%%%%%%%%%%%%%%%%%%
\subsection{Phase damping dynamics}
%%%%%%%%%%%%%%%%%%%%%
Let us first consider a dephasing model, involving only decoherence without dissipation. This type of evolution can be experimentally implemented, for instance, in an optical setup \cite{LLHLGLBP11}, where the environment can be easily tuned so as to produce a variety of dynamics both in the Markovian and non-Markovian regimes. In fact, the time evolution of a photon traveling in a quartz plate may be described by a unitary operator $U(t)$ defined by
\begin{equation}
 U(t)|\lambda\rangle \otimes |\omega\rangle
 = e^{in_{\lambda}\omega t} |\lambda\rangle \otimes |\omega\rangle,
\end{equation}
where $n_{\lambda}$ represents the refraction index for light with polarization $\lambda=H,V$. The presence of the quartz plate thus leads to the decoherence of the superpositions of polarization states, due to the nonzero difference $\Delta n = n_V-n_H$ in the refraction indices of horizontally and vertically polarized photons. 
The corresponding dynamical map $\Lambda_t^*$ in the Schr\"odinger picture takes the form \eqref{evomatrix} with 
\begin{equation}
a(t)=1 \, , \quad c(t) =\kappa(t) \, , 
\end{equation}
and
the decoherence function $\kappa(t)$ is given by the Fourier transform of the frequency distribution of the photon,
\begin{equation}
 \kappa(t) = \int d\omega |f(\omega)|^2 e^{i\omega\Delta n t} \, .
\end{equation}

The frequency distribution is modified in the experiment via variation of the tilting angle of an FP cavity inserted in the path of the photon \cite{LLHLGLBP11}. All such frequency distributions are very well approximated by a sum of two Gaussians centred at frequencies $\omega_k$ with amplitudes $A_k$ and equal widths $\sigma$. This yields
\begin{equation}
\kappa(t)=\frac{1}{1+A}e^{-\frac{1}{2}\sigma^2t^2}(e^{-i\omega_1t}+Ae^{-i\omega_2t})
 \label{kappa}
\end{equation}
where $A_1 = \frac{1}{1+A}$, $A_2 = \frac{A}{1+A}$ and $\Delta\omega=\omega_2-\omega_1$. The tilting angle of the cavity is relatively small and thus the distance $\Delta\omega$ between the peaks remains approximately constant.

%%%%%%%%%%%%%%%%%%%
\subsection{Amplitude damping dynamics}
%%%%%%%%%%%%%%%%%%%

Let us now consider a case where the interaction between the quantum system and its environment also exhibits energy exchange, that is, the interaction is dissipative. We look at a microscopic Hamiltonian model describing a two-state system interacting with a bosonic quantum reservoir at zero temperature given by \cite{BP07}
\eq
H=\omega_o\sigma_z+\sum_k\omega_k a^{\dagger}_k a_k +\sum_k (g_k a_k\sigma_+ +g^*_k a_k^{\dagger}\sigma_-).
\eeq
As usual, $\sigma_{\pm}$ are standard raising and lowering operators respectively. The dynamics of a single amplitude damped qubit is captured by the time-local master equation:
\eq 
\frac{d\varrho_t}{dt}=\gamma(t)\left[\sigma_{-}\varrho_t\sigma_{+}-\frac{1}{2}\{\sigma_{+}\sigma_{-},\varrho_t\}\right],
\eeq
where $\sigma_{\pm}$ are the spin lowering and rising operators. The resulting map $\Lambda_t^*$ can then be written in the form \eqref{evomatrix} with 
\begin{equation}
a(t)=|G(t)|^2 \, , \quad c(t) = G(t) \, ,  
\end{equation}
where
\eq
\gamma(t)=-2\Re\frac{\dot{G}(t)}{G(t)}, \label{gammaADC}
\eeq
the function $G(t)$ depending on the reservoir spectral density. For a qubit interacting resonantly with a leaky cavity, the spectral density has a Lorentzian shape, i.e, 
\begin{align}
J(\omega)=\gamma_M\lambda^2/2\pi[(\omega-\omega_0)^2+\lambda^2]. 
\end{align}
In this case, the function $G(t)$ takes the form 
\eqa
G^L(t)&=&e^{-\lambda t/2}\left[\text{cosh}\left(\sqrt{1-2r}\frac{\lambda t}{2}\right)\right.\nn\\&+&\left.\frac{1}{\sqrt{1-2r}}\text{sinh}\left(\sqrt{1-2r}\frac{\lambda t}{2}\right)\right]
\eeqa
with $r=\gamma_M/\lambda$. For the photonic band bap model (PBG) \cite{PBG}, we have instead
\eqa
G^P(t)&=&2v_1b_1e^{\beta b_1^2+i\Delta_P t}+v_2(b_2+|b_2|)e^{\beta b_2^2 t+i \Delta_P t}\nn\\&-&\sum_{j=1}^{3}v_j|b_j|[1-\Phi(\sqrt{\beta b^2_j t})]e^{\beta b^2_j t+i\Delta_P t}, \label{gPBG}
\eeqa
where $\Delta_P=\tilde\omega_0-\omega_e$ is the detuning from the band gap edge frequency $\omega_e$, set to equal zero as we consider only the resonant case, and $\Phi$ is the error function, whose series and asymptotic representations are given in Ref. \cite{PG}, and 
\eqa
& & b_1=(A_++A_-)e^{i(\pi/4)},\nn\\& & 
b_2=(A_+e^{-i(\pi/6)}-A_-e^{i(\pi/6)})e^{-i(\pi/4),} \nn\\& & 
b_3=(A_+e^{i(\pi/6)}-A_-e^{-i(\pi/6)})e^{i(3\pi/4)},
\eeqa
\eq
A_{\pm}=\left[\frac{1}{2}\pm\frac{1}{2}\left[1+\frac{4}{27}\frac{\Delta_P^3}{\beta^3}\right]^{1/2}\right]^{1/3},
\eeq
\eq
v_1 = \frac{x_1}{(b_1-b_2)(b_1-b_3)} 
\eeq
\eq
v_2 = \frac{b_2}{(b_2-b_1)(b_2-b_3)},
\eeq
\eq
\beta^{3/2}=\tilde\omega_0^{7/2}d^2/6\pi\epsilon_0\hbar c^3.
\eeq
The coefficient $\beta$ is defined as the characteristic frequency, $\epsilon_0$ the Coulomb constant, $d$ the atomic dipole moment, and  $z=\Delta_P/\beta$.

%%%%%%%%%%%%%%%%%%%%
\section{Results and discussion} \label{dyn}
%%%%%%%%%%%%%%%%%%%%

In this section we study the dynamics of incompatibility for the different physically relevant dynamical processes specified in the previous section. 
We are especially interested in comparing the dynamics of incompatibility with that of entanglement, so as to detect the fundamental differences between these quantum resources. In particular, we investigate the expected difference between the timescales of the sudden death of entanglement and incompatibility.

%%%%%%%%%%%%%%%%%%%%%%%%%%%%%%%%%%%%
\subsection{Sudden Death of Incompatibility for Markovian Dynamics}
%%%%%%%%%%%%%%%%%%%%%%%%%%%%%%%%%%%%

\begin{figure}[h]
\includegraphics[width=0.54\textwidth]{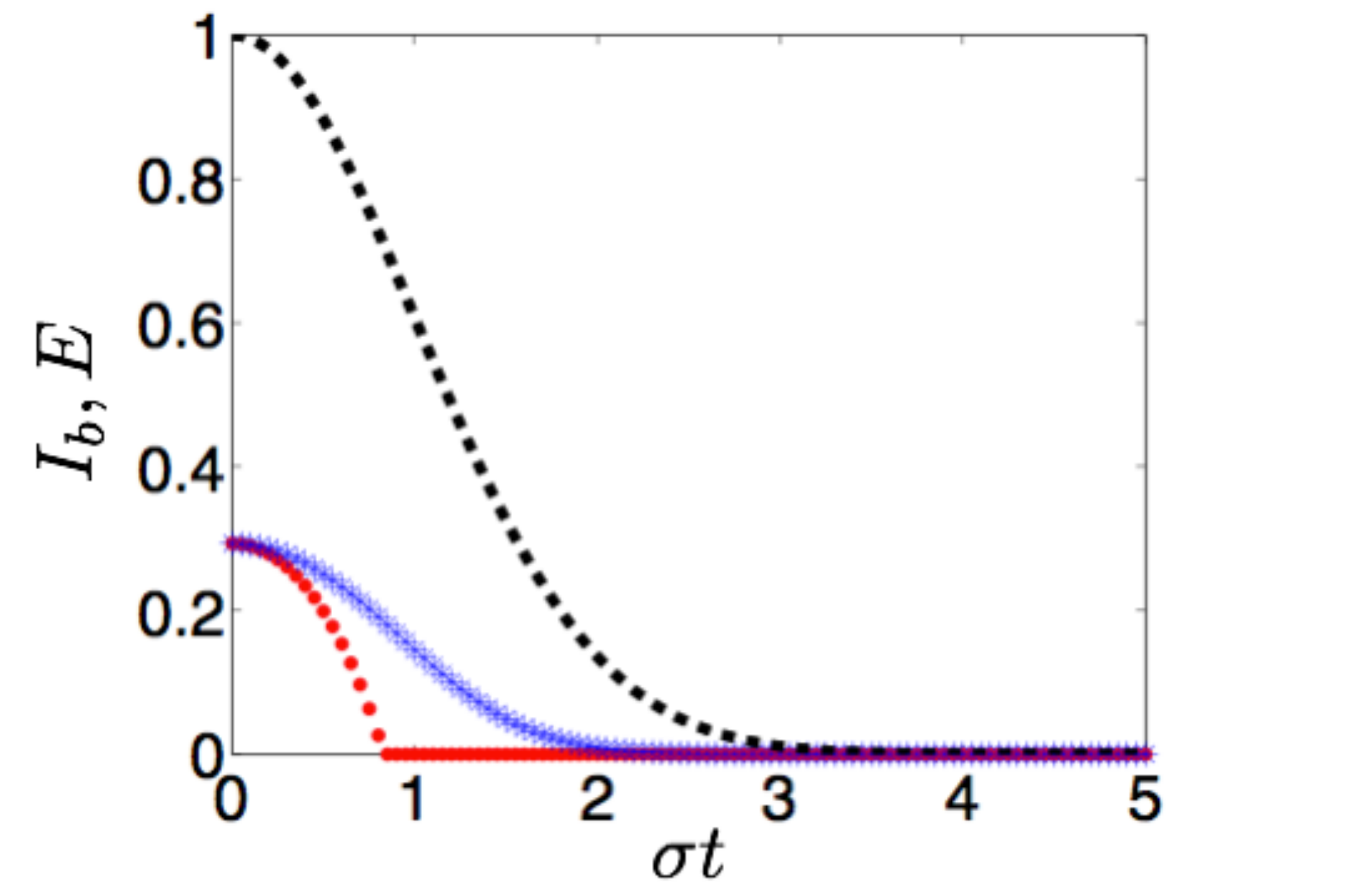}
\caption{(Color online) The evolution of entanglement entanglement in terms of concurrence $\mathcal{E}$ (black squares), and the incompatibility measure $\mathcal{I}_{b=0}$ for the dephasing evolution with Markovian parameters, $A=0$ and $\Delta\omega=2\sigma$. Incompatibility is represented by two initial pairs $\mathcal{P}_1$ (red circles) and $\mathcal{P}_2$ (blue stars).}
\label{figureD1}
\end{figure}

Let us first study the dynamics of incompatibility for the case of Markovian dynamics, using the models presented above. For the models considered here we are in the Markovian regime as long as the entanglement is a monotonically decreasing function. In Fig.~\ref{figureD1} we have plotted the incompatibility for two initial measurement pairs and entanglement of the initial maximally entangled state for the dephasing dynamics. We observe that the incompatibility is extremely fragile under noise. Indeed, entanglement decreases asymptotically towards zero, but incompatibility reaches zero far before the entanglement has vanished. Naturally, for the phase damping evolution, the incompatibility of the measurements in the z-direction (initial pair $\mathcal{P}_2$) is better preserved. The phase damping dynamics does not have a unique fixed point, and hence the the measurement in the z-direction does not evolve towards a trivial measurement. However, the pair $\mathcal{P}_1$ experiences a sudden death of incompatibility at an early stage of the evolution.

%%%%%%%%%%%%%%%%%%%%%%%%%%%%%%%%%%%%%
%%%%%%%%%%%%%%%%%%%%%%%%%%%%%%%%%%%%%

\begin{figure}[h]
\includegraphics[width=0.5\textwidth]{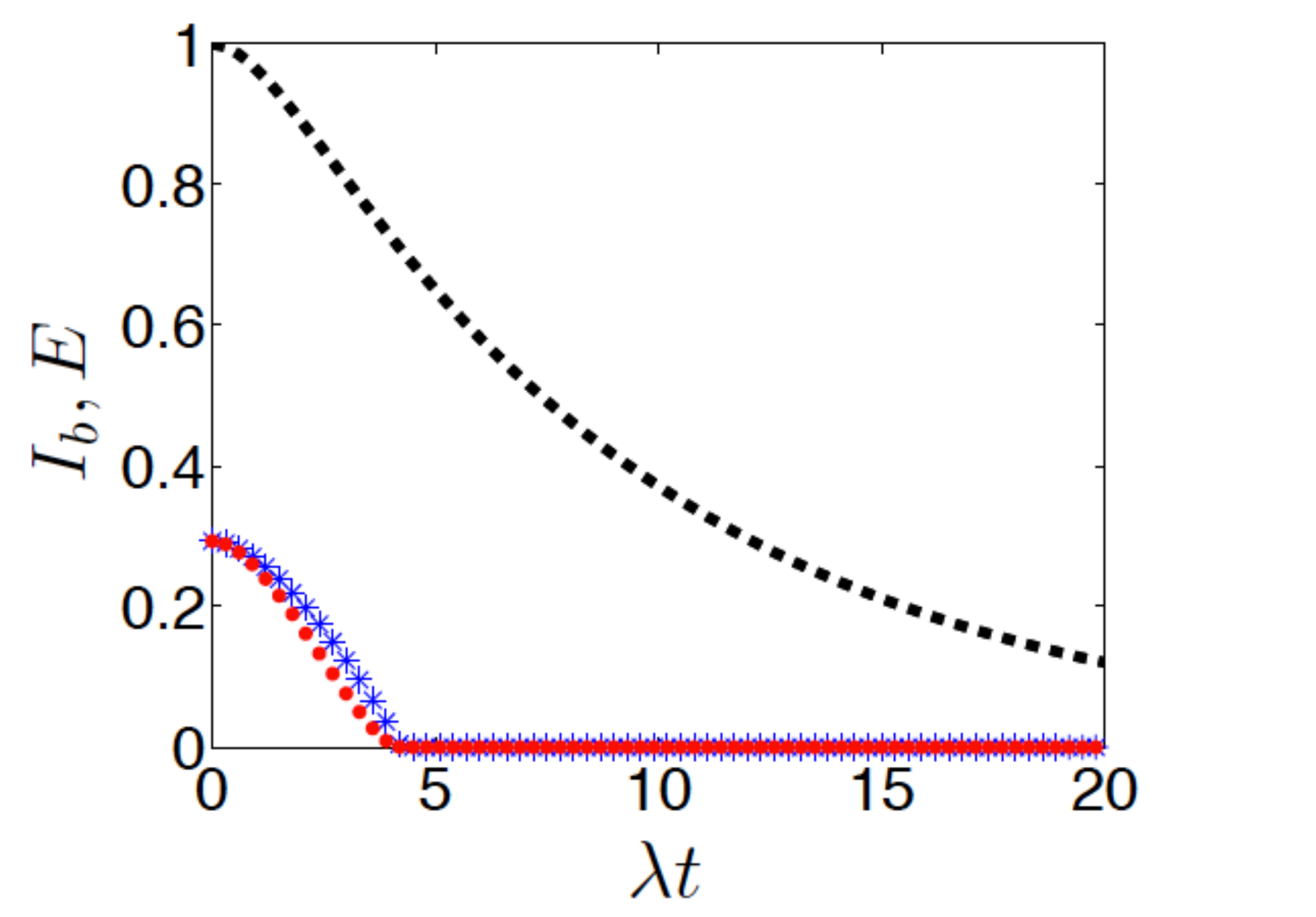}
\caption{(Color online) The evolution of entanglement in terms of concurrence $\mathcal{E}$ (black squares), and the incompatibility measure $\mathcal{I}_{b=0}$ for the amplitude damping dynamics with Lorentzian spectrum and $r=0.2$. Incompatibility is represented by two initial pairs $\mathcal{P}_1$ (red circles) and $\mathcal{P}_2$ (blue stars). }
\label{figureD2}
\end{figure}

In Fig.~\ref{figureD2} we have plotted the incompatibility and concurrence for the Markovian amplitude damping dynamics. We observe the same behaviour as for the dephasing dynamics: a sudden death of incompatibility even with a fairly small amount of noise. For the amplitude damping case, also the initial pair $\mathcal P_2$ exhibits a sudden death of incompatibility.

%%%%%%%%%%%%%%%%%%%%%%%%%%%%%%%%%%%%%
%%%%%%%%%%%%%%%%%%%%%%%%%%%%%%%%%%%%%

%%%%%%%%%%%%%%%%%%%%%%%%%%%%%%%%%%%%%
\subsection{Dynamics of Incompatibility for Non-Markovian Dynamics}
%%%%%%%%%%%%%%%%%%%%%%%%%%%%%%%%%%%%%

As we saw in the previous section, incompatibility is very fragile under Markovian noise. However, quantum features of a system may often be recovered if the environment is tuned such that the dynamics exhibits non-Markovian behaviour \cite{various}. Hence, in this section we study the dynamics of incompatibility under non-Markovian noise, and analyse to what extent the quantumness of measurements can be recovered in this case.

\begin{figure}[!h]
\includegraphics[width=0.53\textwidth]{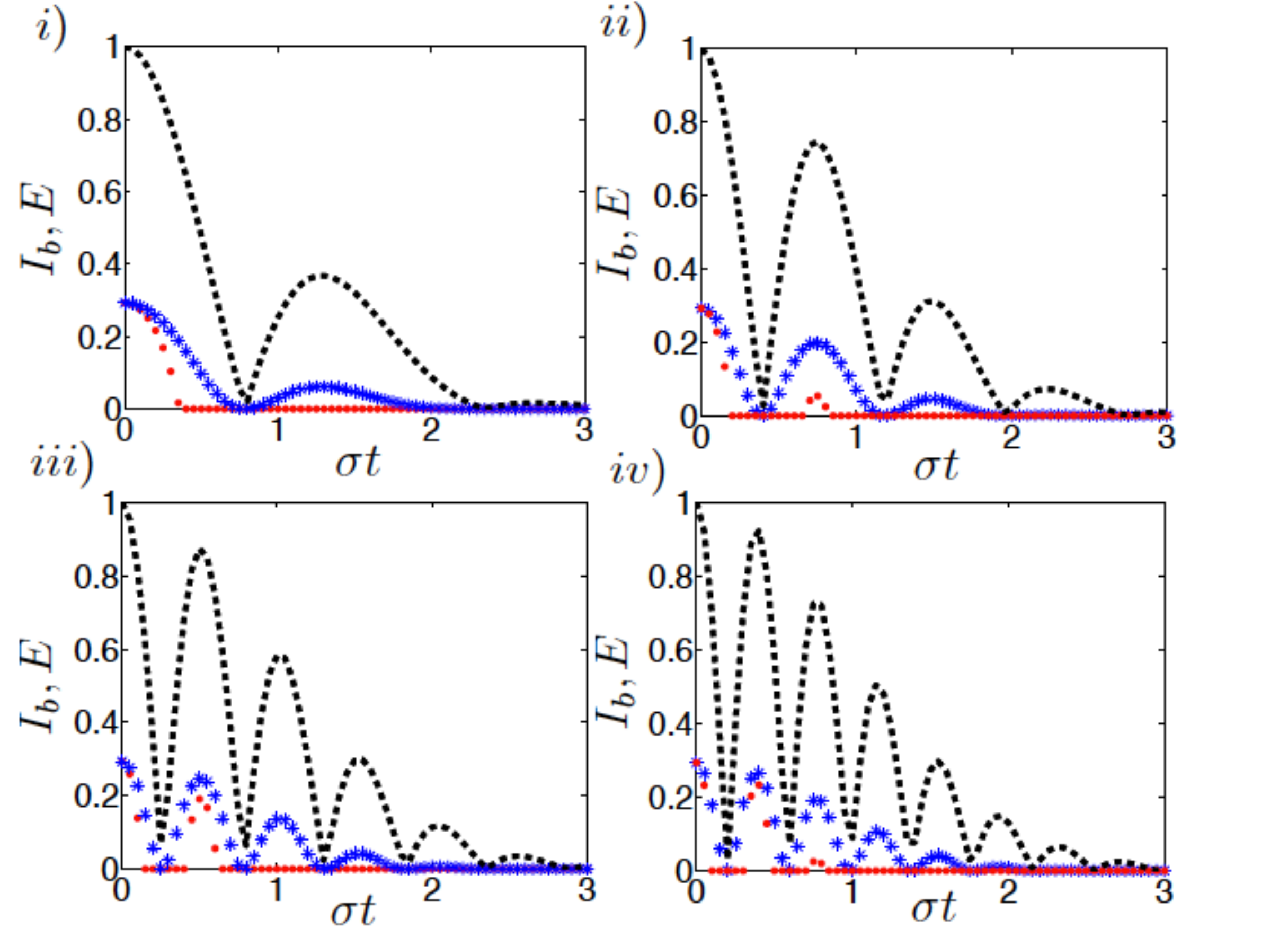}
\caption{(Color online) The evolution of the entanglement in terms of concurrence $\mathcal{E}$ (black squares), and the incompatibility measure $\mathcal{I}_{b=0}$ for the phase damping dynamics with $A=1$ and $\Delta \omega$ equal to i) $4\sigma$, ii) 8$\sigma$, iii) 12$\sigma$ and iv) 16$\sigma$. Incompatibility is represented by two initial pairs $\mathcal{P}_1$ (red circles) and $\mathcal{P}_2$ (blue stars). }
\label{figureD3}
\end{figure}

In Fig.~\ref{figureD3} we have plotted the evolution of the incompatibility and entanglement for the phase damping dynamics in the non-Markovian case. We observe that even with an environment exhibiting considerable non-Markovian character, the incompatibility cannot be very well recovered (Fig.~\ref{figureD3} i)). 
Naturally, for the phase damping case, the measurement pair with the z-direction measurement is again less fragile. We observe that even though a significant portion of entanglement can be recovered, the incompatibility remains absent. When non-Markovianity is increased to the extent that entanglement can be fully recovered, also the incompatibility appears to reach again its maximum value.

%%%%%%%%%%%%%%%%%%%%%%%%%%%%%%%%%%%%%
%%%%%%%%%%%%%%%%%%%%%%%%%%%%%%%%%%%%%

\begin{figure}[!h]
\includegraphics[width=0.53\textwidth]{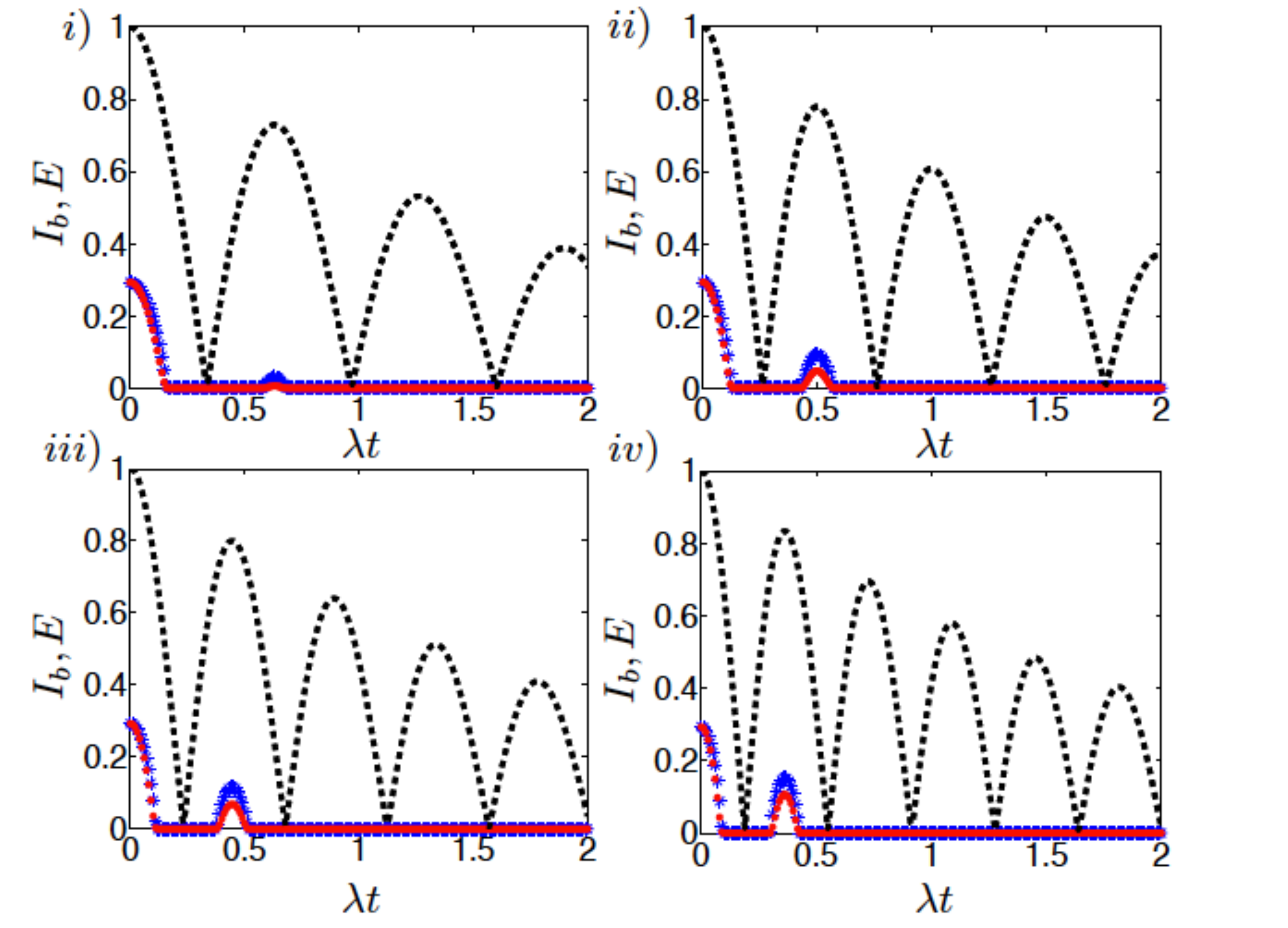}
\caption{(Color online) The evolution of the entanglement in terms of concurrence $\mathcal{E}$ (black squares), and the incompatibility measure $\mathcal{I}_{b=0}$, for the amplitude damping dynamics with Lorentzian spectrum and $r=50$, $80$, $100$ and $150$. Incompatibility is represented by two initial pairs $\mathcal{P}_1$ (red circles) and $\mathcal{P}_2$ (blue stars).}
\label{figureD6}
\end{figure}

For the amplitude damping evolution (Fig.~\ref{figureD6}) we observe that even with highly non-Markovian dynamics, the incompatibility remains zero or at relatively small values, and cannot be fully recovered.

%%%%%%%%%%%%%%%%%%%%%%%%%%%%%%%%%%%%%
%%%%%%%%%%%%%%%%%%%%%%%%%%%%%%%%%%%%%

\begin{figure}[!h]
\includegraphics[width=0.5\textwidth]{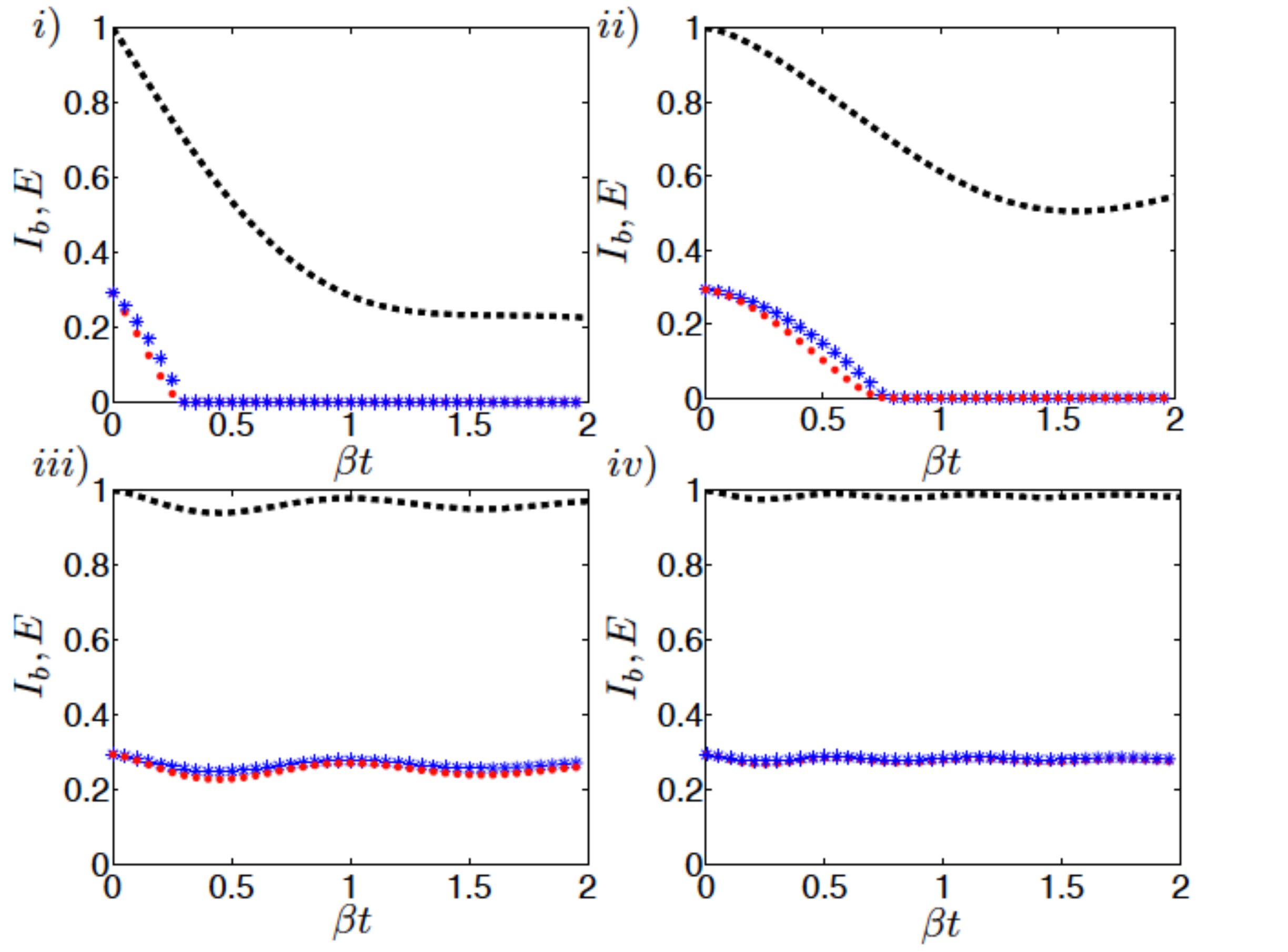}
\caption{(Color online) The evolution of the entanglement in terms of concurrence $\mathcal{E}$ (black squares), and the incompatibility measure $\mathcal{I}_{b=0}$, for the amplitude damped dynamics with PBG spectral density and $z=1$, $0$, $-5$ and $-10$. Incompatiblity is represented the two initial pairs $\mathcal{P}_1$ (red circles) and $\mathcal{P}_2$ (blue stars).}
\label{figureD7}
\end{figure}

We then proceed to consider a highly engineered environment, where a photonics band gap (PBG) tunes the dynamics such that as the non-Markovianity increases, and the initial entanglement can be very well protected. In Fig.~(\ref{figureD7}) we plot the entanglement and incompatibility dynamics for this model, observing that even in the case where entanglement can be well protected  (Fig.~\ref{figureD7} ii)) the incompatibility experiences a sudden death without recovery. Naturally, if the entanglement is fully maintained also the incompatibility persists.

%%%%%%%%%%%%%%%%%%%%%%%%%%%%%%%%%%%%%
%%%%%%%%%%%%%%%%%%%%%%%%%%%%%%

%%%%%%%%%%%%%
\section{Conclusions} \label{conclusions}
%%%%%%%%%%%%%%%%%%%%%%%%%%%%%%%%%%%%%
Since the discovery that quantum phenomena can be harnessed as resources for computation and communication, a considerable effort has been made towards understanding how to protect quantum systems from the effects of noise. The basic understanding is that quantum properties tend to get destroyed as the system interacts with an environment, and hence extensive quantitative studies on the behaviour of quantum resources such as entanglement and discord under noisy dynamics have been undertaken. However, a definite understanding of the essence of quantum resources remains elusive, as the relevant quantifications crucially depend on the application at hand. In contrast to the usual viewpoint where quantum resources are associated exclusively to states, we emphasise that non-classicality may also be attributed to measurements used to extract information on the quantum system. This becomes especially relevant in practical scenarios where the set of available measurements is strongly limited by experimental constraints. Thus, it is crucial to study how the quantum properties of the measurements are influenced by noise.

In this paper, we investigated the evolution of incompatibility of a pair of quantum measurements under two commonly studied dynamical maps, dephasing (in the $z$-direction) and amplitude damping. While these hardly exhaust the collection of physically relevant two-level dynamics, they already represent different basic aspects of lossy quantum dynamics, with dephasing consisting of pure decoherence, and amplitude damping including also dissipation. More general treatment will be a topic of future work.

Our starting point was the general result on incompatibility breaking channels \cite{ibc}, which says that sudden death of initially maximal entanglement always implies sudden death of \emph{all} incompatibility, while the converse is not true. We investigated this phenomenon in the specific physically relevant settings. In particular, we found that, indeed, even in the absence of sudden death of entanglement, a sudden death of incompatibility often occurs. Incompatibility is known to be a crucial resource for many quantum protocols for which entanglement is not sufficient. Thus, it is possible that even in the absence of entanglement sudden death the relevant quantum resource may have been already destroyed.

We also studied how non-Markovian dynamics allows one to recover incompatibility. We found that in many cases where a significant ``portion" of entanglement can be recovered, incompatibility cannot. Thus, non-Markovian dynamics may not be able to recover all quantum resources, even if entanglement is recovered. We believe that this study will be useful for developing a more resource-oriented view of non-Markovian dynamics.

In contrast to concurrence and other entanglement measures, our choice of the incompatibility measure depends on the initial measurements; we demonstrated the dependence by using two different initial measurement pairs. This basis dependence is actually motivated by the fact that the relevant applications mentioned above (i.e. CHSH-Bell and steering scenarios) specifically involve a restricted selection of measurements. In this sense, evolution of incompatibility should be seen as a \emph{purpose-oriented} view to decoherence. For instance, we could quantify noise-robustness of a steering-based key distribution protocol by incompatibility. Concerning our specific choice of the incompatibility measure \cite{ourpaper}, there are alternatives, such as ``incompatibility weight" \cite{Pusey15} based on the steering quantification in \cite{Sk14}. The resulting incompatibility measure has recently been compared with ours in \cite{UoBuGuPe14} (see especially Fig. 2 therein); while there are strong quantitative differences, the qualitative behaviour is relatively similar except for special cases.

Finally, we wish to point out that our work on the dynamics of incompatibility in open systems has a natural application in quantum optimal control, namely finding optimal measurement resources for, say, noisy steering by using the incompatibility quantification as a figure of merit. Such a study has recently been carried out by one of the present authors \cite{KB15}.

\section*{Acknowledgements}
S.M., E.-M.L. and C.A. acknowledge funding from the EU Horizon 2020 research and innovation programme under grant agreement No 641277, the Magnus Ehrnrooth Foundation, and the Academy of Finland (Project no. 287750). J.K. acknowledges support from the EPSRC (projects EP/J009776/1 and EP/M01634X/1).
%%%%%%%%%%%%%


\newpage

\begin{thebibliography}{10} 

\bibitem{Horo09} R. Horodecki, R., P. Horodecki, M. Horodecki, K. Horodecki, {\em Rev. Mod. Phys.} \textbf{81} 865 (2009).

\bibitem{BrGo15} F.G.S.L. Brandao, G. Gour, arXiv:1502.03149 [quant-ph].

\bibitem{CoFrSp14} B. Coecke, T. Fritz, R.W. Spekkens, arXiv:1409.5531 [quant-ph].

\bibitem{We89} R.F. Werner, {\em Phys. Rev. A} \textbf{40} 4277 (1989).

\bibitem{WoPeFe09}
M.M.~Wolf, D.~Perez-Garcia, C.~Fernandez, {\em Phys.~Rev.~Lett.} \textbf{103} 230402 (2009).

\bibitem{QuVeBr14} M.T. Quintino, T. V\'ertesi, N. Brunner, {\em Phys. Rev. Lett.} \textbf{113} 160402 (2014).

\bibitem{Sk14} P. Skrzypczyk, M. Navascu\'es, D. Cavalcanti,  {\em Phys. Rev. Lett.} \textbf{112} 180404 (2014).

\bibitem{UoMoGu14} R. Uola, T. Moroder, O. G\"uhne, {\em Phys. Rev. Lett.} \textbf{113} 160403 (2014).
\bibitem{UoBuGuPe14} R. Uola, C. Budroni, O. G\"uhne, J.-P. Pellonp\"a\"a, {\em Phys. Rev. Lett.} \textbf{113} 160403 (2014).

\bibitem{WiJoDo07} H.M. Wiseman, S.J. Jones, A.C. Doherty, {\em Phys. Rev. Lett.} \textbf{98} 140402 (2007).

\bibitem{BP07} H.-P. Breuer, F. Petruccione, {\em The Theory of Open Quantum Systems}, Oxford University Press, 2007.

\bibitem{Da76} E.B. Davies, {\em Quantum Theory of Open Systems}, Academic Press, 1976.

\bibitem{ESD1} T. Yu and J. H. Eberly, {\em Science} \textbf{323}, 598 (2009).

\bibitem{TimeIND} P. Haikka, T. H. Johnson, and S. Maniscalco {\em Phys. Rev. A} \textbf{87}, 010103(R) (2013).

\bibitem{ibc}
T.~Heinosaari, J.~Kiukas, D.~Reitzner, J. Schultz, {\em J. Phys. A: Math. Theor.} \textbf{48} 435301 (2015).

\bibitem{PhaseD} J. Luczka, {\em Physica A} {\bf 167}, 919 (1990); G. M. Palma, K.-A. Suominen, and A. K. Ekert, {\em Proc. R. Soc. London, Ser. A} {\bf452}, 567 (1996); J. H. Reina, L. Quiroga and N. F. Johnson, {\em Phys. Rev. A} {\bf65}, 032326 (2002).

\bibitem{ENTtrap} B. Bellomo, R. Lo Franco, S. Maniscalco, and G. Compagno, {\em Phys. Rev. A} \textbf{78}, 060302(R) (2008).

\bibitem{Pusey15} M.F. Pusey, {\em J. Opt. Soc. Am. B} \textbf{32} A56 (2015).

\bibitem{ourpaper}
T.~Heinosaari, J.~Kiukas, D.~Reitzner, {\em Phys. Rev. A} \textbf{92} 022115 (2015).

\bibitem{GrPoWi05} B. Groisman, S. Popescu, A. Winter, {\em Phys. Rev. A} \textbf{72} 032317 (2005).

\bibitem{erkka} E. Haapasalo,  {\em J. Phys. A: Math. Theor.} \textbf{48} (2015) 255303.

\bibitem{AlPiBaToAc07} M. L. Almeida, S. Pironio, J. Barrett, G. T\'oth, A. Ac\'in, {\em Phys. Rev. Lett.} \textbf{99} 040403 (2007).

\bibitem{BaGaGhKa13} M. Banik, Md.Rajjak Gazi, S. Ghosh, G. Kar, {\em Phys. Rev. A} \textbf{87} 052125 (2013).


\bibitem{BuHeScSt13} P. Busch, T. Heinosaari, J. Schultz, and N. Stevens, {\em Europhys. Lett.} \textbf{103} 10002 (2013).

\bibitem{Busch}
P.~Busch, H.-J.~Schmidt, {\em Quant.~Info.~Proc.} \textbf{9} 143 (2010).

\bibitem{SRH08} 
P.~Stano, D.~Reitzner, T.~Heinosaari, {\em Phys.~Rev.~A} \textbf{78} 012315 (2008).

\bibitem{Yu10} 
S.~Yu, N.-L.~Liu, L.~Li, C.H.~Oh, {\em Phys.~Rev.~A} \textbf{81} 062116 (2010).

\bibitem{GRTZ02} N. Gisin, G. Ribordy, W. Tittel, H. Zbinden, {\em Rev. Mod. Phys.} \textbf{74} 145 (2002).

\bibitem{PiWa14} M. Piani, J. Watrous, {\em Phys. Rev. Lett.} \textbf{114} 060404 (2015).

\bibitem{Wootters98} W.K. Wootters, {\em Phys. Rev. Lett.} \textbf{80} 2245 (1998).

\bibitem{NMPRL} H.P. Breuer, E.-M. Laine, J. Piilo, {\em Phys. Rev. Lett.} \textbf{103} 210401 (2009).

\bibitem{Wang} J. Wang, H. Batelaan, J. Podany, A.F. Starace {\em J. Phys. B: At. Mol. Opt. Phys.} \textbf{39} (2006) 4343--4353.

\bibitem{various} S. Maniscalco, Stefano Olivares, and Matteo G. A. Paris, {\em Phys. Rev. A} \textbf{75}, 062119 (2007); R. Vasile, P. Giorda, S. Olivares, M. G. A. Paris, and S. Maniscalco, {\em Phys. Rev. A} \textbf{ 82}, 012313 (2010); S. McEndoo, P. Haikka, G. De Chiara, G. M. Palma and S. Maniscalco, {\em Europhys. Lett.} \textbf{101}, 60005 (2013); C. Addis, P. Haikka, S. McEndoo, C. Macchiavello, and S. Maniscalco, {\em Phys. Rev. A} \textbf{87}, 052109 (2013); C. Benedetti, M.G.A. Paris, and S. Maniscalco, {\em Phys. Rev. A} \textbf{89}, 012114 (2014); B. Bylicka, D. Chruscinski, and S. Maniscalco, {\em Scientific Reports} \textbf{4}, 5720 (2014); 

\bibitem{LLHLGLBP11} B.-H. Liu, L. Li, Y.-F. Huang, C.-F. Li, G.-C. Guo, E.-M. Laine, H.-P. Breuer, J. Piilo, {\em Nature Physics} \textbf{7} 931 (2011).

\bibitem{PBG} S. John and T. Quang, Phys. Rev. A {\bf 50}, 1764 (1994).

\bibitem{PG} I. S. Gradshteyn and I. M. Ryzhik, Table Integral, Series,and Products (Academic, New York, 1980), p. 931.

\bibitem{KB15} J. Kiukas, D. Burgarth, arXiv:1509.08822 (2015).

\end{thebibliography}
\end{document}